\documentclass[a4paper,fleqn]{cas-dc}

\usepackage[numbers]{natbib}
\usepackage{amssymb}
\usepackage{amsmath}
\usepackage{xcolor}
\usepackage{amsthm}
\usepackage{caption}
\usepackage{float}
\usepackage{subcaption}
\def\tsc#1{\csdef{#1}{\textsc{\lowercase{#1}}\xspace}}
\tsc{WGM}
\tsc{QE}
\begin{document}
\let\WriteBookmarks\relax
\def\floatpagepagefraction{1}
\def\textpagefraction{.001}

\shorttitle{}    

\shortauthors{}  

\title [mode = title] {Generation of High-order Laguerre–Gaussian modes in Coated and Uncoated Graded-Index and Step-Index Multimode Fibers}

\author[1]{Wasyhun Asefa Gemechu}[type=editor,
                        % auid=000,bioid=1,
                        % prefix=Sir,
                        % role=Researcher,
                        ]
%\fnref{fn1}}
\credit{Writing - original draft, Methodology, Investigation, Formal analysis, Data curation, Conceptualization}
\ead{wasyhun.gemechu@unibs.it}
\author[1]{Umberto Minoni}[%
% role=Co-ordinator,
% suffix=Jr,
]
%\ead{umberto.minoni@unibs.it}
\author[1]{Michela Borghetti}[%
% role=Co-ordinator,
% suffix=Jr,
]
%\ead{michela.borghetti@unibs.it}
\author[2]{Fabrizio Frezza}[%
% role=Co-ordinator,
% suffix=Jr,
]
\author[1]{Daniele Modotto}[%
% role=Co-ordinator,
% suffix=Jr,
]
\credit{Writing – review $\&$ editing, Supervision, Project administration}
%\ead{daniele.modotto@unibs.it}
\author[2]{Stefan Wabnitz}[%
% role=Co-ordinator,
% suffix=Jr,
]
\author[3]{Fabio Mangini}[%
% role=Co-ordinator,
% suffix=Jr,
]
\credit{Writing – review $\&$ editing, Project administration,  Conceptualization}
\cortext[cor1]{Corresponding author}
%\fntext[fn1]{This is the first author footnote.}
%\fntext[fn2]{Another author footnote.}
%\fntext[fn3]{Yet another author footnote.}
\affiliation[1]{organization={Dipartimento di Ingegneria dell’Informazione, Università di Brescia},
addressline={via Branze 38},
postcode={25123},
city={Brescia},
country={Italy},
}

\affiliation[2]{organization={Dipartimento di Ingegneria dell’Informazione, Elettronica e Telecomunicazioni, Sapienza Università di Roma},
addressline={via Eudossiana 18},
postcode={00184},
city={Rome},
country={Italy},
}

\affiliation[3]{organization={Dipartimento di Ingegneria, Università degli Studi Niccolò Cusano},
addressline={Via don Carlo Gnocchi 3},
postcode={00166},
city={Rome},
country={Italy},
}



\begin{abstract}
We report an efficient experimental method for generating high-order Laguerre–Gauss (HOLG) modes by simply coupling a Gaussian beam into the cladding of a multimode fiber (MMF). In particular, the order of the HOLG mode remains invariant with respect to input power, propagation distance, and pulse duration. Furthermore, spectral and power measurements confirm that the beam-shaping mechanism is predominantly linear, whereas Kerr nonlinearity primarily affects the longitudinal phase-matching condition and conversion efficiency, without altering the generated mode order. Altogether, these findings establish our approach as a highly robust and scalable platform for generating tailored optical beams.
\end{abstract}


% \begin{highlights}
% \item 
% \item 
% \item 
% \end{highlights}

%\nocite{*}

% Keywords
%% Keywords

\maketitle

\section{Introduction} \label{intro}
The ability to shape light in space has become one of the defining capabilities of modern photonics. Increasingly, advances arise not from the development of new laser sources themselves but from controlling how light is distributed after it is generated—its phase, intensity profile, and modal structure \cite{rubinsztein2017roadmap}. Among structured optical fields, the Laguerre–Gaussian (LG) modes occupy a particularly important place. Due to their distinctive dependence on the angular phase, $ \exp(i \ell \phi)$, which gives photon orbital angular momentum (OAM) while preserving modal orthogonality \cite{allen1992orbital}, these beams are physically distinct and technologically significant. High-order LG (HOLG) modes are now indispensable across multiple domains. In the field of optical micromanipulation, OAM carrying beams are utilized to rotate microscopic particles, biological cells, and micro-machines, functioning much like an optical spanner \cite{padgett2011tweezers, grier2003revolution, he1995direct, forbes2021structured}. These OAM-carrying beams are more stable than traditional Gaussian traps. In the realm of optical communications, orthogonal spatial modes provide an additional dimension for multiplexing, such as space-division multiplexing (SDM), helping to overcome the capacity limits of single-mode fiber networks \cite{bozinovic2013terabit, liu20221, richardson2013space, pang2018400, olaleye2023generation}. Additionally, since quantum light technologies require access to a vast, multidimensional state space, OAM modes are being used to encode more information, distribute entanglement, and facilitate quantum sensing \cite{molina2007twisted, erhard2020advances, sit2017high, forbes2024quantum}. Beyond these established areas, structured beams are increasingly used in nonlinear optics to modify phase-matching conditions \cite{wright2015controllable}, in advanced microscopy for contrast enhancement and super-resolution imaging \cite{pushkina2021superresolution, hu2023experimental, ma2021high, iketaki2012super}, and in laser materials processing, where vortex beams influence energy deposition and ablation morphology \cite{hnatovsky2010materials, tan2022simultaneous}. 

Across various fields, reliably and efficiently generating higher-order spatial modes without introducing excessive optical complexity remains a challenge. Most state-of-the-art techniques still depend on free-space beam shaping. Although spatial light modulators (SLMs) \cite{forbes2016creation, ohtake2007universal}, spiral phase plates \cite{ruffato2014generation}, q-plates \cite{vertchenko2017laguerre, marrucci2006optical, marrucci2011spin}, and metasurfaces \cite{yue2016vector, devlin2017arbitrary} can produce a diverse range of structured beams, they often come with drawbacks such as high insertion loss, wavelength sensitivity, limited optical damage thresholds, and alignment constraints that complicate integration into compact or fiber-based systems \cite{ohtake2007universal}. Although vortex beams can be generated directly through laser-cavity engineering, these methods often sacrifice tunability or add complexity to the overall system \cite{forbes2019structured,scholes2020general}. 

Due to their optical confinement, all-fiber-based solutions are gaining popularity, as they inherently offer mechanical stability, compactness, and the ability to handle high optical power \cite{tran2023exploration}. Due to their robust intermodal coupling and the diverse modal landscape, multimode fibers (MMFs) present an attractive option for structural light generation. Stable propagation of the OAM beam has been demonstrated in specialized fibers, such as ring-core and photonic-crystal fibers \cite{tandje2019ring}. In addition, certain modes can be selectively excited using photonic lanterns or long-period gratings \cite{wu2017all, ramachandran2002band}. However, these methods are often limited by narrowband resonant structures or necessitate custom fabrication. In contrast, standard graded-index (GRIN) multimode fibers offer a simpler alternative. Their parabolic index profile makes them excellent candidates for passive spatial-mode transformation, as it minimizes intermodal dispersion, promotes periodic self-imaging, and accommodates modal families that closely resemble LG modes \cite{wright2015controllable}. Using resonant coupling between the cladding and the core modes, we recently reported a grating-free method for passive Gaussian-to-HOLG mode conversion in a conventional GRIN multimode fiber \cite{Gemechu2024, gemechu2025direct}. The resulting beams exhibited stable multi-ring profiles during fiber propagation as the input polarization and optical power of the Gaussian input beam were varied. These findings suggest that the resonant coupling between the cladding and core modal manifolds serves as an intrinsic beam-shaping mechanism, enabling the fiber to perform spatial transformations effectively \cite{hansson2020nonlinear, krupa2016observation}.

Despite advances in the field, many important questions remain. In particular, there has been no thorough investigation of how propagation length affects modal purity. Furthermore, more information is required to fully understand how temporal dynamics—especially pulse duration and temporal walk-off—affect conversion efficiency. The impact of the fiber coating is another frequently overlooked element; polymer coatings can selectively attenuate cladding modes, thereby altering the pool of available modes for resonant coupling and affecting stability and efficiency. The present study systematically investigates both GRIN and step-index (SI) MMFs, with and without protective coatings, to address the open questions outlined above. Through a systematic investigation of the excitation pulse duration, coating conditions, and propagation length, we hope to establish the operational limits for alignment-free, passive production of HOLG beams in conventional MMFs.

\section{Experimental Setup} \label{setup}
The experimental setup follows the scheme reported in \cite{gemechu2025direct}, with adjustments tailored to the present study. A Yb-based pulsed laser (Light Conversion PHAROS-SP-HP) operating at 1030 nm served as the light source, with pulse durations ($\tau$) tunable between $0.5$ and $8~ps$ and repetition rates ranging from 2 to 100 kHz. The power and polarization state of the input beam, characterized by a near-Gaussian profile ($M^2 \approx 1.3$), were adjusted using a variable neutral-density filter (VDF) and a half-wave plate (HWP), respectively. The laser beam was focused onto the fiber input facet using a lens $\text{L}_1$ with a focal-length of $f=50~mm$, resulting in a waist diameter of approximately $15~\mu m$. Two types of MMFs were investigated: a GRIN (Thorlabs GIF50E, $\text{NA}=0.200\pm 0.015$) and SI (Thorlabs FG050LGA, $\text{NA}=0.22\pm0.02$) having a core and cladding diameter of $50~\mu m$ and $125~\mu m$, respectively. This combination allowed us to directly compare the influence of the refractive-index profile on the Gaussian-to-HOLG beam transformation. To uncover the subtle yet crucial role of the polymer coating, we performed comparative measurements on both coated and uncoated fibers.

We performed a cutback measurement starting from a 2-m-long fiber, removing 20 cm segments sequentially while monitoring the output power and near-field spatial profiles. To ensure unperturbed launch conditions throughout the cut-back process, the fiber segment closest to the input facet was rigidly secured with tape and magnets to prevent unintended fiber movement. Under centered-core excitation, the coupling efficiency reached approximately $82\%$, confirming high-quality fiber facets and providing a reliable benchmark for off-axis cladding launch used in the LG transformation experiments. At the fiber output, the near-field spatial beam profile was recorded using a CCD beam profiler (Gentec-EO Beamage-3.0). The output beam was collimated using a telescope configuration ($\text{L}_2:~f = 2.75~mm$ and $\text{L}_3:~f = 400~mm$). Spectral purity was ensured by a narrow bandpass filter (BPF) centered at 1030 nm (with FWHM: $5~nm$). A flip-mirror (FM) enabled seamless transitions between spatial profiling and power or spectral measurements using an optical power meter (PM, Thorlabs PM16-122) and an optical spectrum analyzer (OSA, Yokogawa AQ6370D). 

\begin{figure}[pos=h]%% placement specifier
\centering
\includegraphics[width=1.0\linewidth]{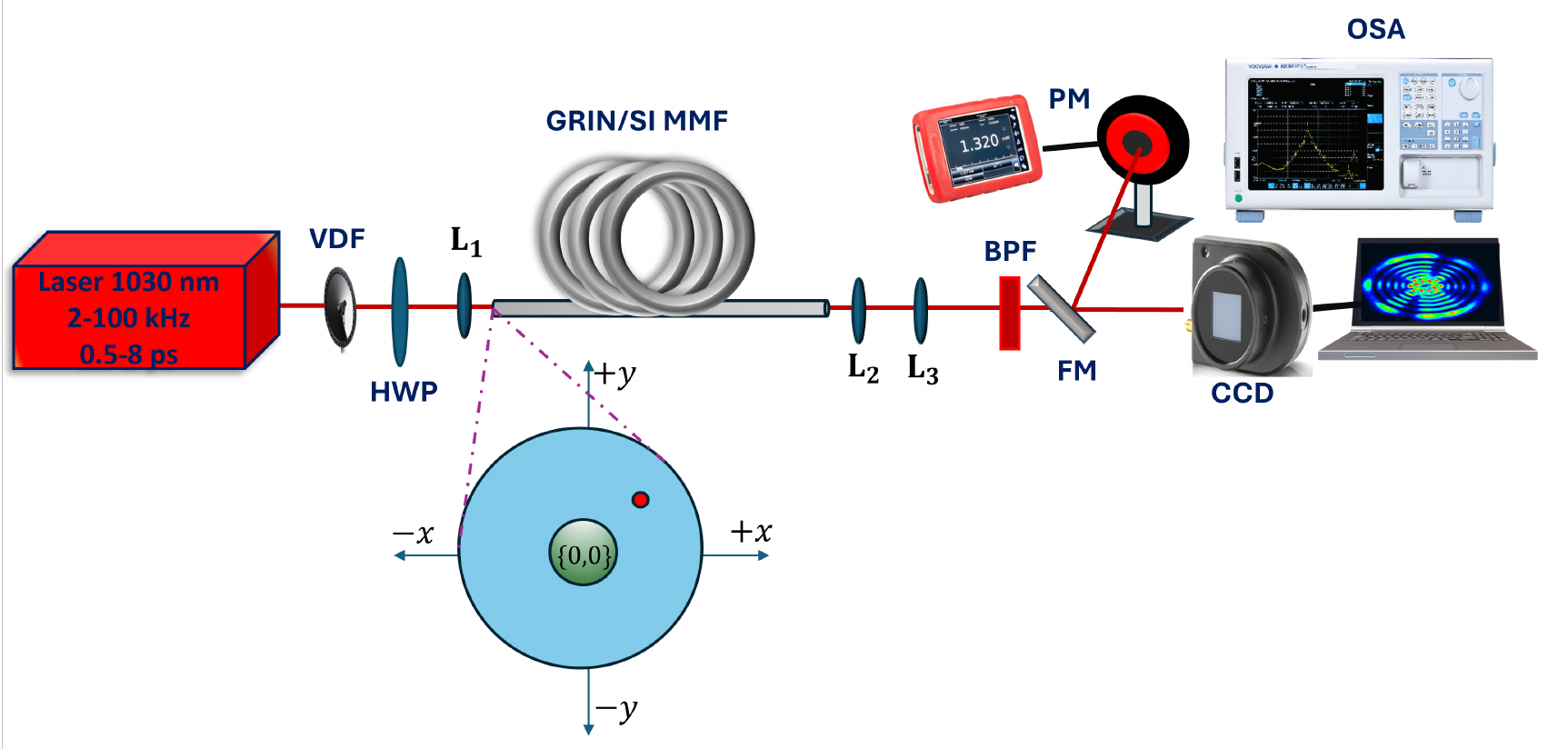}
\caption{Experimental setup for generating HOLG mode in MMFs.The components include: VDF (Variable neutral density filter), HWP (Half-wave plate), $\text{L}_{1-3}$ (Lenses), SI/GRIN MMF (Step-index/Graded-index multimode fiber), BPF (Band-pass filter), FM (Flipping Mirror), PM (Power Meter), and OSA (Optical Spectrum Analyzer). The inset details the injection conditions and the $\pm\{x,y\}$ offsets between the fiber axis and the input laser beam.}
\label{fig:setup}
\end{figure}

\section{Experimental results: GRIN MMFs}
\begin{figure}[pos=h]
    \centering
    \includegraphics[width=1.0\linewidth]{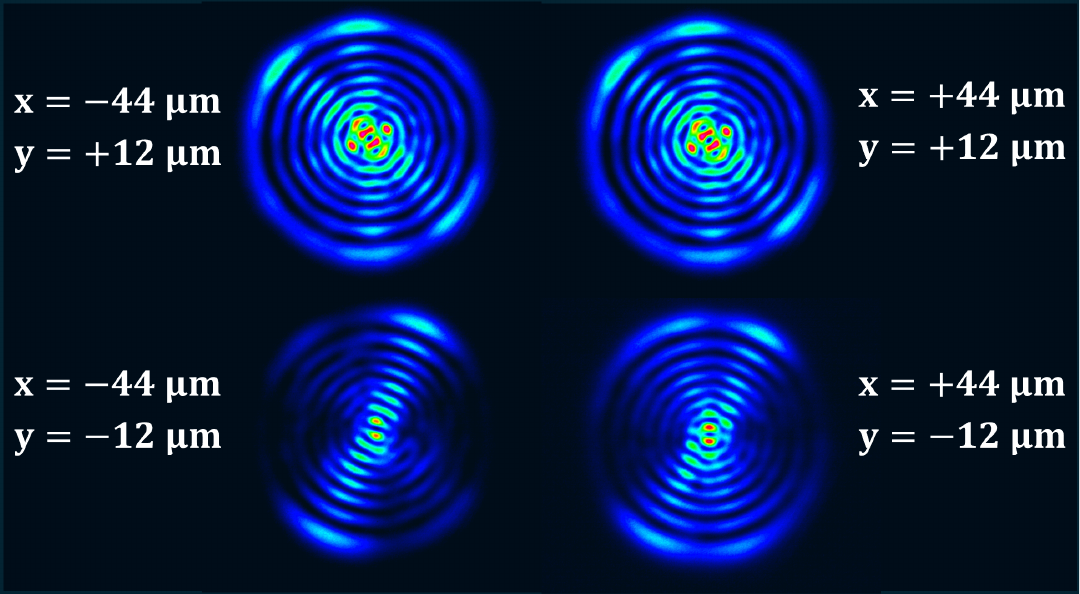}
    \caption{The experimentally generated HOLG modes arise from selected symmetrical offset values in both the x- and y-directions within the GRIN MMF. These HOLG modes are formed in the core, with the edge of the intensity profile corresponding to the core-cladding interface.}
    \label{fig:Symmetry}
\end{figure}

The HOLG beams were generated following the spatial-offset coupling technique described in \cite{gemechu2025direct}. The process began by on-axis coupling of the input Gaussian laser beam at the center of the fiber (coordinate $\{0,0\}$, Fig.~\ref{fig:setup} inset) to excite the fundamental mode. Using this as a reference point, we introduced incremental transverse offsets along both the x ($\pm x$) and y ($\pm y$) axes. Real-time near-field monitoring with a CCD camera captured the evolution of the output from a speckled distribution to highly structured light at a specific spatial offset (e.g., $x=+44~\mu m$, $y=+12~\mu m$, indicated by the red marker). This transition signifies resonant coupling between the fundamental cladding mode and the the $14^{\text{th}}$ core-mode group, resulting in the generation of an $\text{LG}_{7,0}$ mode \cite{gemechu2025direct, ivanov2006cladding}. The excitation of the $\text{LG}_{7,0}$ mode is a direct consequence of the specific refractive index profile of the GRIN MMF. This parabolic profile ensures that the propagation constants of the cladding and core modes satisfy the phase-matching condition, $\Delta \beta \approx 0$ \cite{gemechu2025direct}. We further characterized the robustness of this beam-shaping method by examining its symmetry properties. Given that the fundamental cladding mode is an intensity-null distributed across the entire cladding region (see Fig. 4 in \cite{gemechu2025direct}), introducing identical offsets in the x and y directions relative to the fiber axis should produce equivalent power transfer to the core. Experimental results in Fig.~\ref{fig:Symmetry} confirm this: HOLG mode of the same order—characterized by an invariant number of concentric rings—are consistently generated by adjusting the shift values along either axis (i.e., $x = \pm 44~\mu m$ and $y = \pm 12~\mu m$). This symmetry demonstrates that the observed beam shaping is not a stochastic occurrence but a deterministic, controllable transition. The coupling is sustained by the natural intersection of the dispersion curves for the graded-index core and the acrylic/air-silica cladding \cite{gemechu2025direct}.   

\begin{figure}[pos=h]
    \centering
    \includegraphics[width=1.0\linewidth]{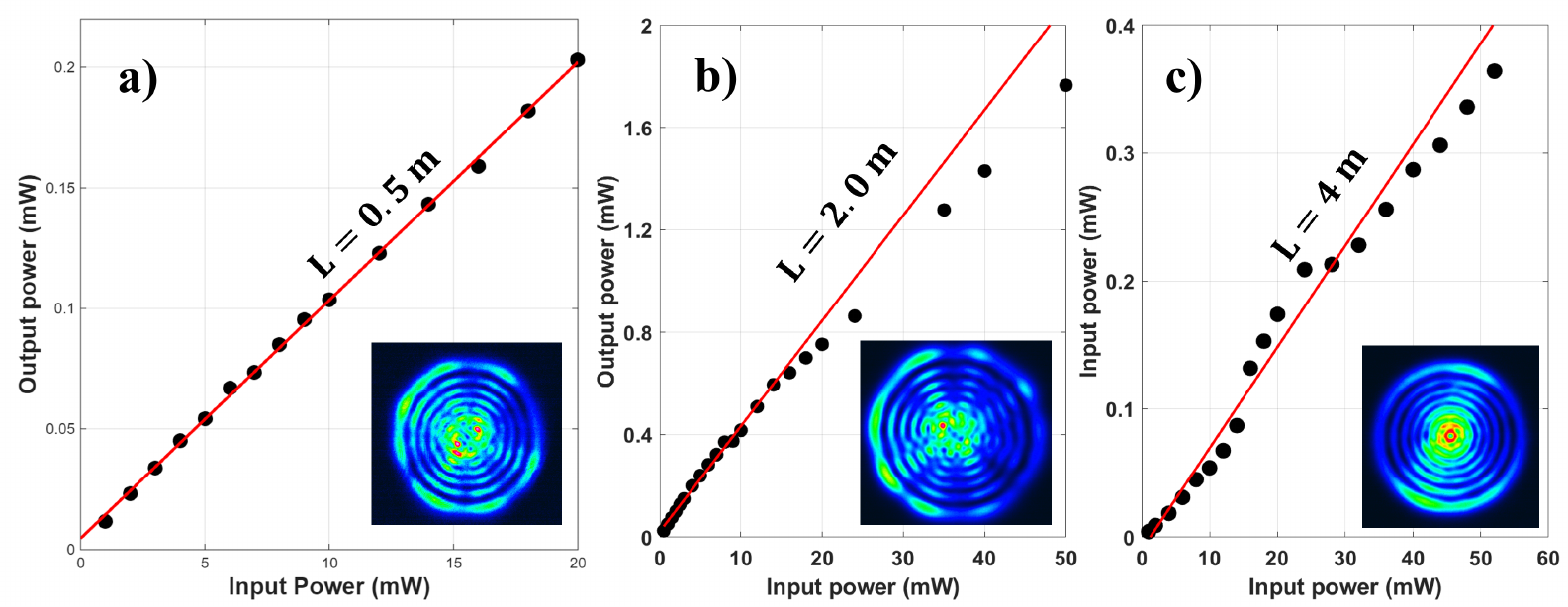}
    \caption{Experimentally measured relationship between input and output power (black dots) as a function of fiber length for the GRIN MMF at different lengths: (a) $\text{L} = 0.5$ m, (b) $\text{L} = 2.0$ m, and (c) $\text{L} = 4.0$ m. The linear fit is indicated by the red line.}
    \label{fig:Trans_vs_len}
\end{figure}

To rigorously validate the robustness of our HOLG beam generation, we initially investigated the beam behavior over varying propagation distances. By comparing fiber segments ranging from $L=0.5~\text{m}$ to $L=4.0~\text{m}$, as shown in Fig.~\ref{fig:Trans_vs_len}, we monitored both the generation efficiency and the spatial intensity of the output. Remarkably, despite an eight-fold increase in fiber length, the near-field intensity distributions (refer to Fig.~\ref{fig:Trans_vs_len}, insets) maintain a nearly identical azimuthal symmetry and radial mode order (number of characteristic concentric rings). Quantitatively, the overall transmissivity (the ratio of output to input power) remains consistent across these lengths, ranging from $1$ to $4\%$. Fig.~\ref{fig:Trans_vs_len} also demonstrates that, for average input powers below $14$ mW, the conversion process remains strictly linear across most lengths, with the shortest fiber maintaining linearity throughout its range. This linear regime confirms that the underlying generation mechanism is a passive resonant tunneling process, in which the mode-coupling coefficient depends solely on modal overlap rather than on the input beam intensity \cite{ivanov2006cladding}. However, once the input power exceeds $14$ mW, the system exhibits nonlinear behavior. For example, in a 2-meter-long fiber, we observe a monotonic decrease in transmissivity, dropping from a peak of $1.27\%$ to $0.88\%$ at higher powers (Fig.~\ref{fig:Trans_vs_len}b). In a 4-meter segment, this high-power regime causes the transmissivity to oscillate between $0.43\%$ and $0.87\%$ (Fig.~\ref{fig:Trans_vs_len}c). This deviation can be attributed to the onset of the optical Kerr effect, in which nonlinear effects primarily affect the longitudinal dynamics (coupling efficiency), while the transverse modal order remains unchanged \cite{ivanov2006cladding}.

\begin{figure}[pos=h]
    \centering
    \includegraphics[width=1.0\linewidth]{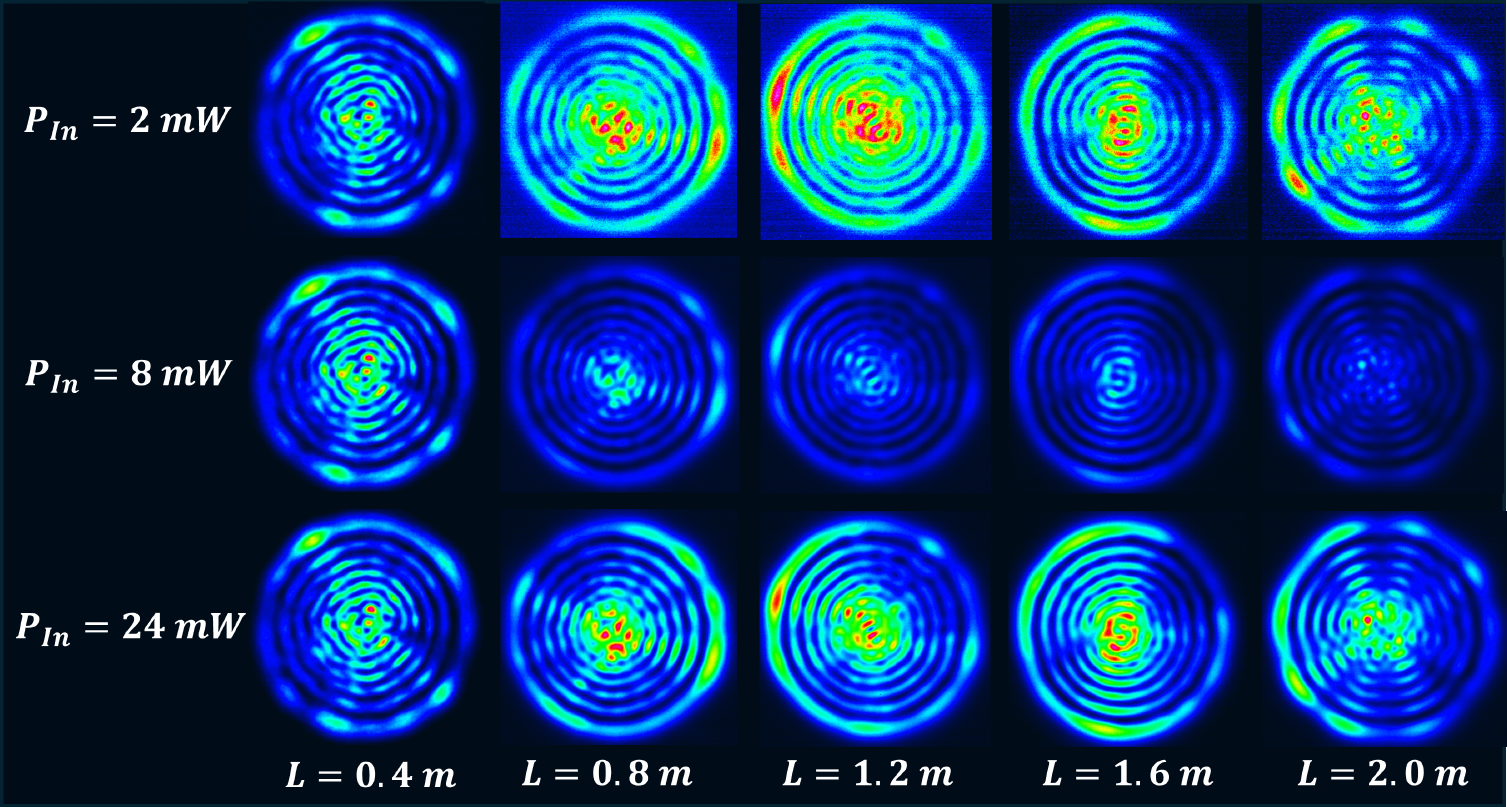}
    \caption{Stability of generated HOLG beams. Transverse near-field intensity distributions measured as a function of input power and propagation length ($0.4 \leq \text{L} \leq 2.0$~m). }
    \label{fig:int_pwr_len}
\end{figure}

To visualize the beam shaping's resilience to these linear and nonlinear propagation dynamics, we conducted a systematic cutback experiment, capturing the near-field intensity profiles as the fiber was shortened from $\text{L} = 2.0$~m to $\text{L} = 0.4$~m. As shown in Fig.~\ref{fig:int_pwr_len}, comparing these transverse mode profiles across different input powers and propagation lengths reveals a striking topological invariance. The fundamental signature of the HOLG beam—the exact number of concentric rings—remains locked in place across the entire parameter space. Interestingly, a minor spatial asymmetry emerges at the 2-meter mark, manifesting as a slight discontinuity within the circular intensity fringes. Yet, even with this localized perturbation, the overarching radial structure and the total ring count survive completely intact. This visual evidence compellingly demonstrates that once the structured light field is generated, its spatial topology is remarkably robust, weathering macroscopic changes in both fiber length and input power without losing its modal identity.

\begin{figure}[pos=h]
    \centering
    \includegraphics[width=1.0\linewidth]{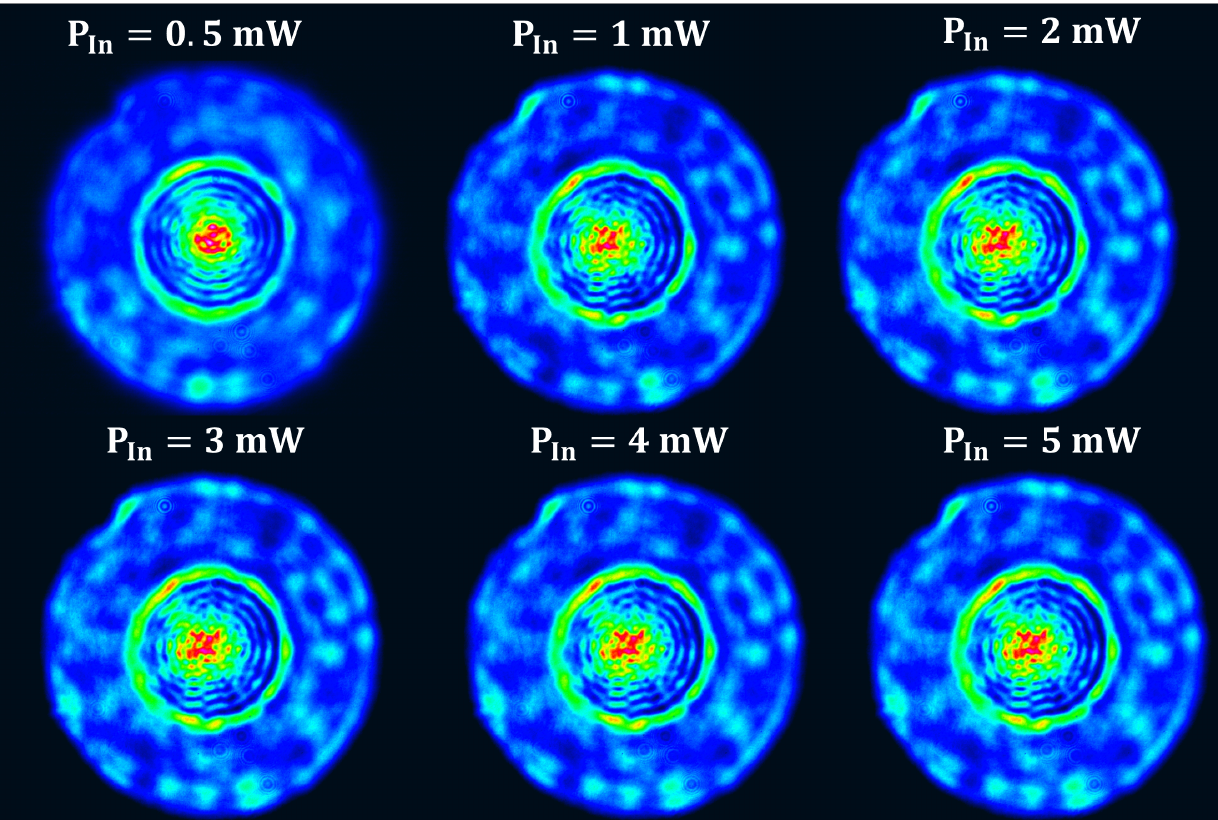}
    \caption{Near-field intensity distribution in an uncoated GRIN fiber. The bare cladding-air interface prevents the absorption of cladding modes, making both the generated HOLG beam (core) and the residual pump light (cladding) distinctly visible. The blue ring immediately preceding the intense green ring marks the boundary between the core and the cladding. }
    \label{fig:uncoated}
\end{figure}

To isolate the specific boundary conditions governing resonant coupling, we conducted a complementary experiment in which the high-index acrylic coating was removed, and the bare silica surface was cleaned with acetone to remove residual polymer and contaminants. We then replicated our HOLG beam-generation protocol and captured the near-field intensity profile of a 1-meter-long fiber. Fig.~\ref{fig:uncoated} illustrates the output of this uncoated fiber, demonstrating that the HOLG modes remain well-formed and strictly confined within the core. Notably, as the input power was increased from 0.5 mW to 5 mW, the radial mode order of the generated beam remained invariant, highlighting the exceptional spatial stability of the resonant coupling process. However, in contrast to the coated fiber—where the surrounding area appears dark—the core HOLG mode in the bare fiber is surrounded by a prominent, intricately structured light field that fills the entire cross-section of the cladding. This visual distinction is directly influenced by the physics of cladding modes and their boundary conditions, as detailed in \cite{ivanov2006cladding}. In a coated fiber, the protective polymer coating typically has a refractive index higher than that of the pure silica cladding and exhibits strong absorption. As a result, any light coupled into the cladding behaves as a leaky mode; the polymer serves as a sink, rapidly draining and attenuating the cladding energy through radiation and absorption losses. However, removing this coating replaces the silica-polymer boundary with a silica-air interface. Given that the refractive index of air ($\text{n}_{\text{Air}} \approx 1.0$) is significantly lower than that of silica ($\text{n}_{\text{Silica}} \approx 1.45$), this new boundary creates a substantial index contrast ($\Delta n \approx 0.45$). The previously leaky cladding modes undergo total internal reflection at the bare glass-air boundary, transforming into strictly guided, strongly bound modes. Consequently, the propagation loss of the cladding-mode in the uncoated fiber is significantly lower than that of the coated fiber. Comparing near-field images from both coated and uncoated fibers, the cladding mode in the case of the coated fiber is attenuated due to polymer absorption within the initial few centimeters, indicating that the stable HOLG mode we observe at the output is primarily generated within the initial few centimeters of the fiber, as suggested by the absence of visible cladding light at the output. In stark contrast, the uncoated fiber allows for continuous, cumulative resonant tunneling of energy from the cladding into the core along its entire length.

\begin{figure}[pos=h]
    \centering
    \includegraphics[width=1.0\linewidth]{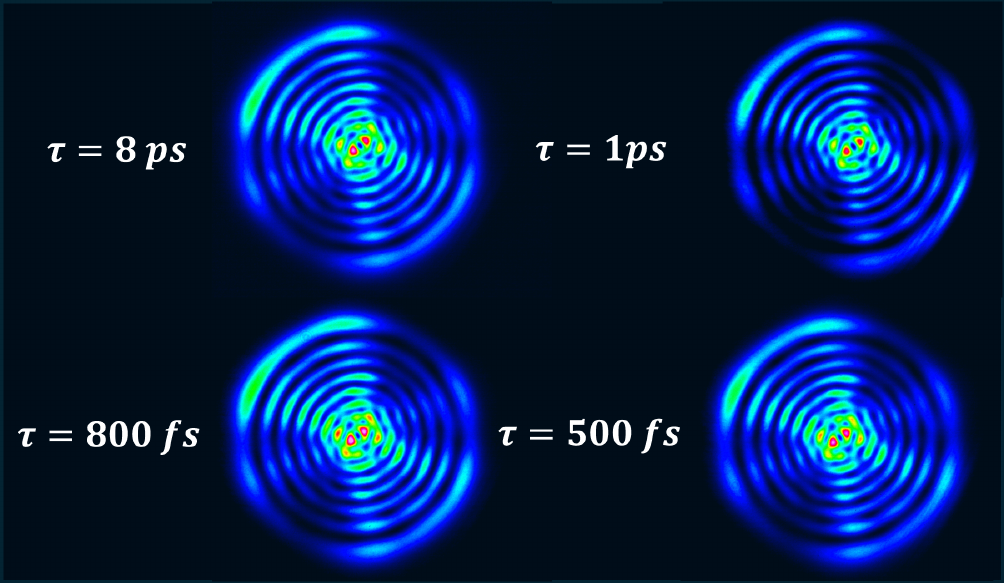}
    \caption{Near-field image of HOLG mode at an input power of 24 mW using a Gaussian pulse of different durations in a 1-meter-long GRIN MMF.}
    \label{fig:walkoff}
\end{figure}

Having demonstrated the robustness of our Gaussian beam-shaping method—where precise phase-matching enables resonant tunneling of energy from the cladding to the core—we must now explore the implications of extending this mechanism to the ultrafast-pulse regime. In this context, a critical temporal constraint arises: the walk-off effect. This phenomenon is a direct consequence of the group velocity mismatch (GVM) between the interacting modes. Due to the different dispersion landscapes experienced by the core modes (guided by a parabolic refractive index profile) and the cladding modes (determined by the silica-air or silica-acrylic boundary), these modes propagate at significantly different group velocities \cite{gemechu2025direct}. As an ultrafast pulse traverses the medium, the GVM causes the corresponding wave packets to diverge rapidly in time. Such temporal separation drastically reduces longitudinal mode overlap, thereby truncating the effective interaction length and leading to a notable decrease in energy-transfer efficiency. Using a simplified estimate based on the fiber NA and representative group-index values, we obtain an effective walk-off parameter $\Delta \beta_1 =-51$ fs/mm, corresponding to a temporal walk-off length $L_W = \tau/|\Delta \beta_1|$ \cite{gemechu2025direct}. We emphasize that the relevant quantity is the modal-group-index mismatch between the specific phase-matched cladding and core modes; therefore, $L_W$ should be interpreted as an order-of-magnitude estimate of the effective interaction length. For an input pulse duration of $\tau=8$ ps, the interaction length is approximately $L_W \approx 14.7$ cm. In contrast, for a sub-picosecond pulse with $\tau=0.5$ ps (500 fs), $L_W$ reduces significantly to just $9.8$ mm. To examine the impact of temporal walk-off on our beam-shaping mechanism, we launched both picosecond and sub-picosecond pulses into a 1-meter-long coated GRIN multimode fiber. By systematically tuning the pulse duration to 8 ps, 1 ps, 0.8 ps, and 0.5 ps and adjusting the input power from 1 mW to 50 mW, we mapped the system's spatial response. Near-field profiles, collected at an input power of 24 mW (Fig.~\ref{fig:walkoff}), reveal that the HOLG modes clearly maintain their ring structure and angular symmetry, with their mode order remaining entirely invariant. This invariance is a direct result of the decoupling of space and time in the beam-shaping process. The walk-off effect strictly limits the longitudinal resonance-coupling efficiency, thereby reducing the total power transferred to the core. However, these longitudinal dynamics are fundamentally orthogonal to the transverse spatial evolution of the HOLG modes. This spatial invariance is rigorously supported by the coupled-mode theory of optical fibers, as detailed by Ivanov et al. \cite{ivanov2006cladding}. According to this framework, the energy transfer between cladding and core states is strictly governed by the transverse spatial overlap integral of their respective electric fields. Because the HOLG modes are rigid, orthogonal spatial eigenmodes mathematically defined by the core's parabolic index profile, this overlap integral evaluates to zero for any non-symmetric cladding perturbations. Consequently, the core boundary acts as a highly selective spatial filter; it necessitates that once phase-matched resonant coupling is achieved within the first few centimeters of the fiber, any tunneled energy is immediately projected into the strict, discrete eigen-basis of the core. Thus, the topological signature of the field—defined by its invariant radial nodes ($p$) and conserved orbital angular momentum ($\ell$)—remains permanently locked by the waveguide geometry, propagating unperturbed despite the severe longitudinal temporal walk-off. 

\begin{figure}[pos=h]
\centering
\begin{subfigure}{0.49\textwidth}
\includegraphics[width=0.95\linewidth, height=5cm]{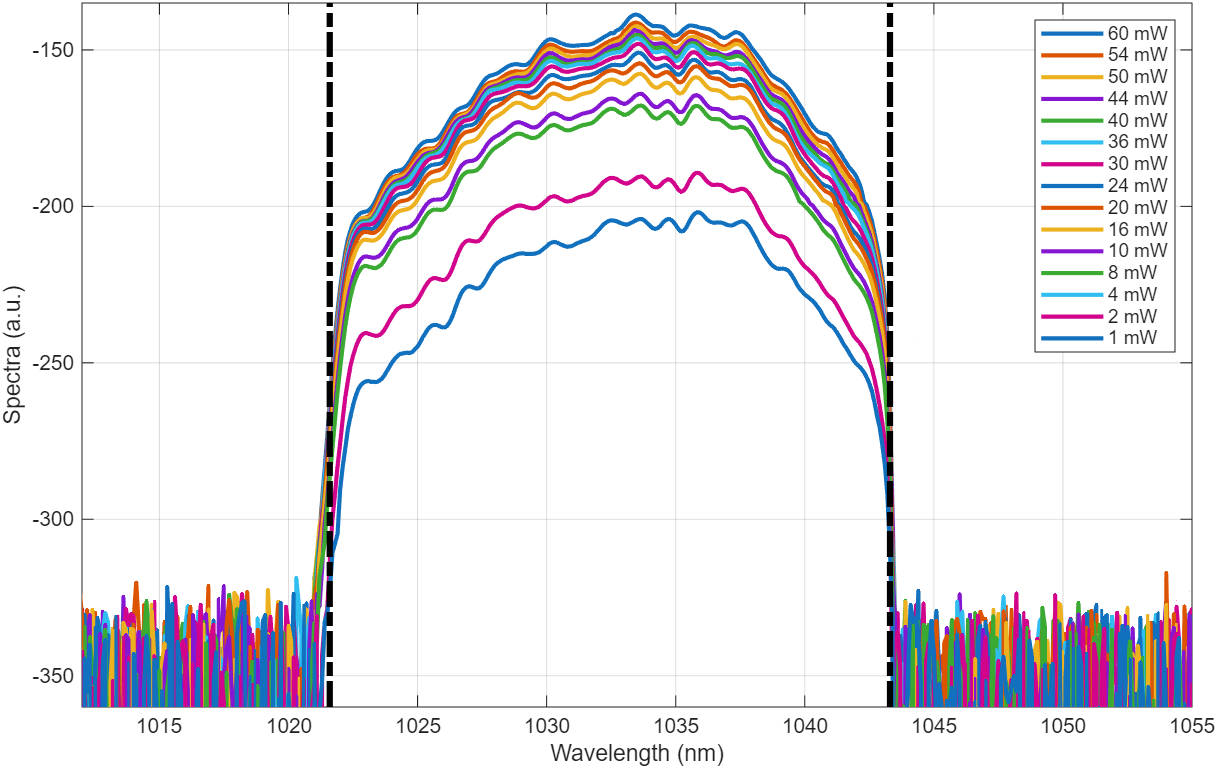} 
\caption{}
\label{fig:spec_8ps}
\end{subfigure}
\begin{subfigure}{0.49\textwidth}
\includegraphics[width=0.95\linewidth, height=5cm]{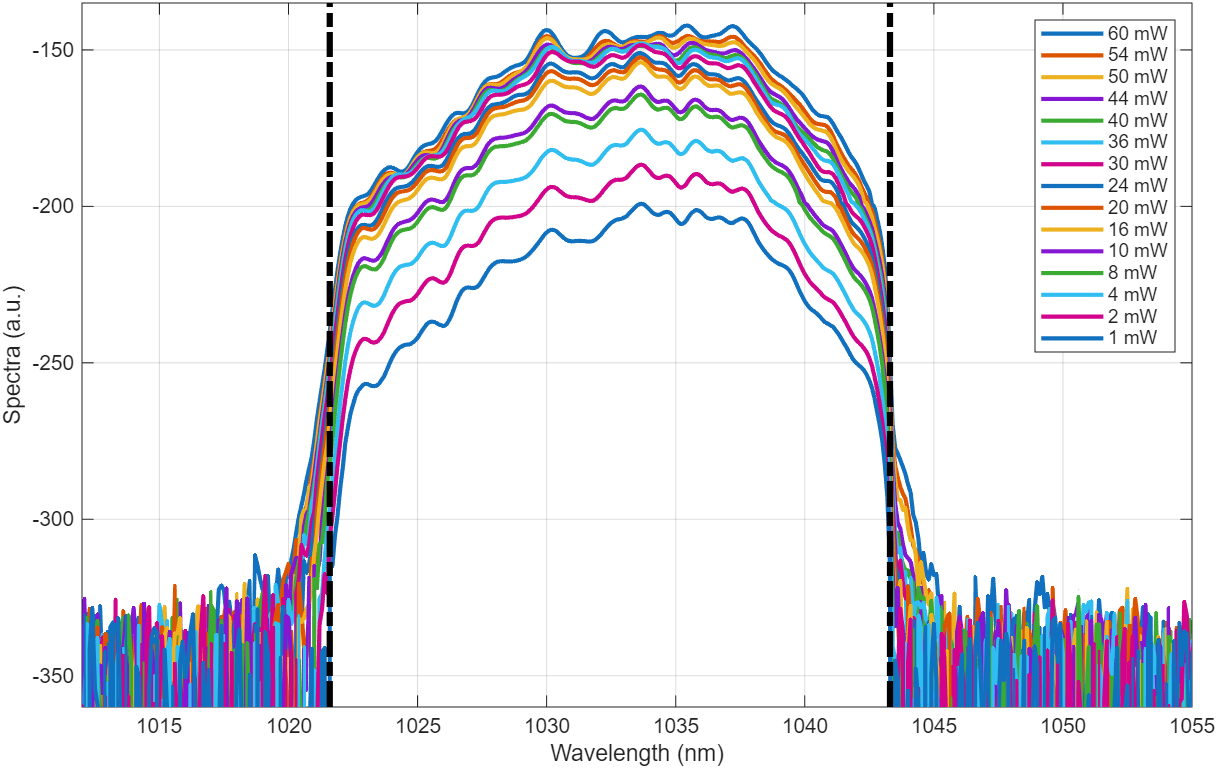} 
\caption{}
\label{fig:spec_1ps}
\end{subfigure}
\begin{subfigure}{0.49\textwidth}
\includegraphics[width=0.95\linewidth, height=5cm]{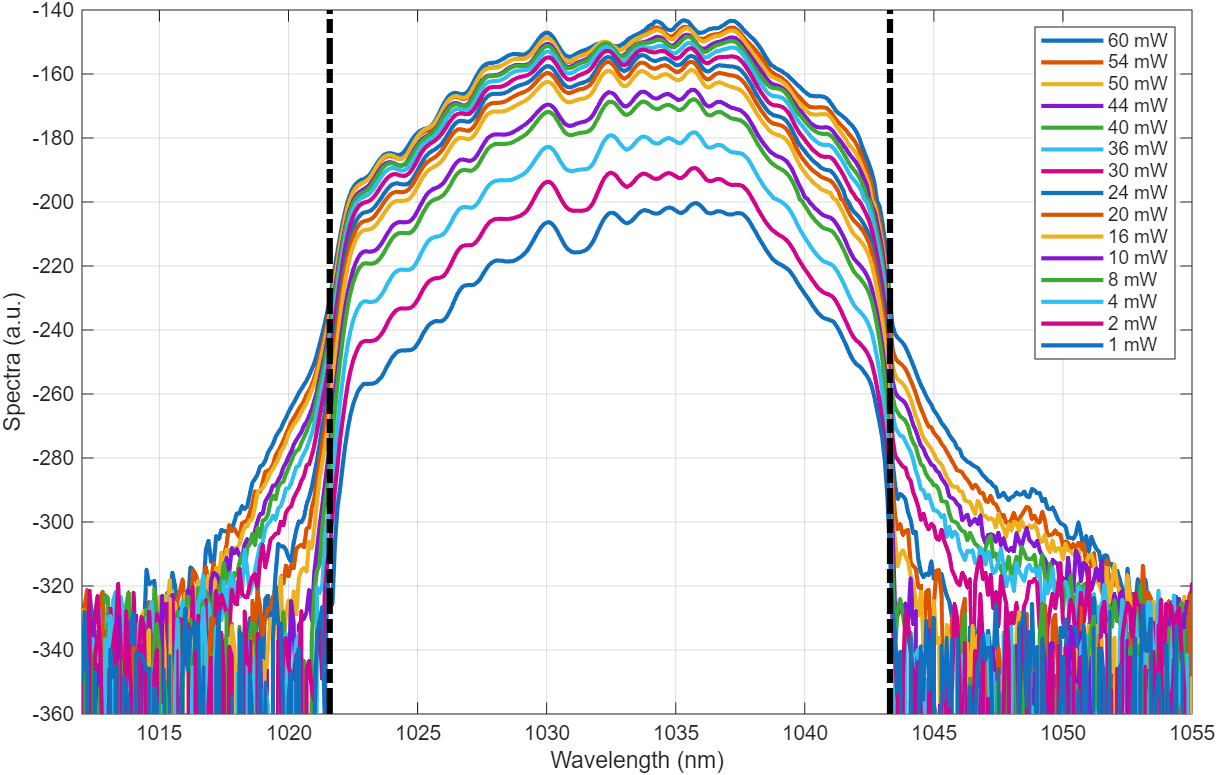}
\caption{}
\label{fig:spec_500fs}
\end{subfigure}
\caption{Spectral output measured after a 1-meter-long fiber as a function of input average power for different input pulse durations (a) $\tau = 8$~ps, (b) $\tau=1$~ps, and (c) $\tau = 0.5$~ps. The black-dashed lines at wavelengths 1022 and 1043 nm serve as visual guides. }
\label{fig:image2}
\end{figure}

To confirm that the beam-shaping mechanism is governed by a linear optical process, we measured the spectral profile of the output beam as a function of the average input power, ranging from 1 mW to 60 mW, for three pulse widths: 0.5 ps, 1 ps, and 8 ps. We connect the output fiber using APC connectors before inserting it into the OSA input. The resulting output spectra exhibited a clear dependence on the duration of the pulse, which is directly correlated with the power transmissivity variation illustrated in Fig.~\ref{fig:Trans_vs_len}. For the relatively longer pulse duration ($\tau = 8$~ps), the spectrum remained essentially unchanged across all power levels tested in a 1-meter-long fiber (Fig.~\ref{fig:spec_8ps}). In contrast, when the pulse duration was compressed to the sub-picosecond regime ($\tau = 0.5$~ps), characterized by high peak power, a noticeable spectral broadening was observed at high power levels (see Fig.~\ref{fig:spec_500fs}). This broadening is a classic signature of Self-Phase Modulation (SPM) driven by the intrinsic Kerr nonlinearity of the material($n_2$).

\section{Experimental results: SI MMFs}
\begin{figure}[pos=h]
    \centering
    \includegraphics[width=1.0\linewidth]{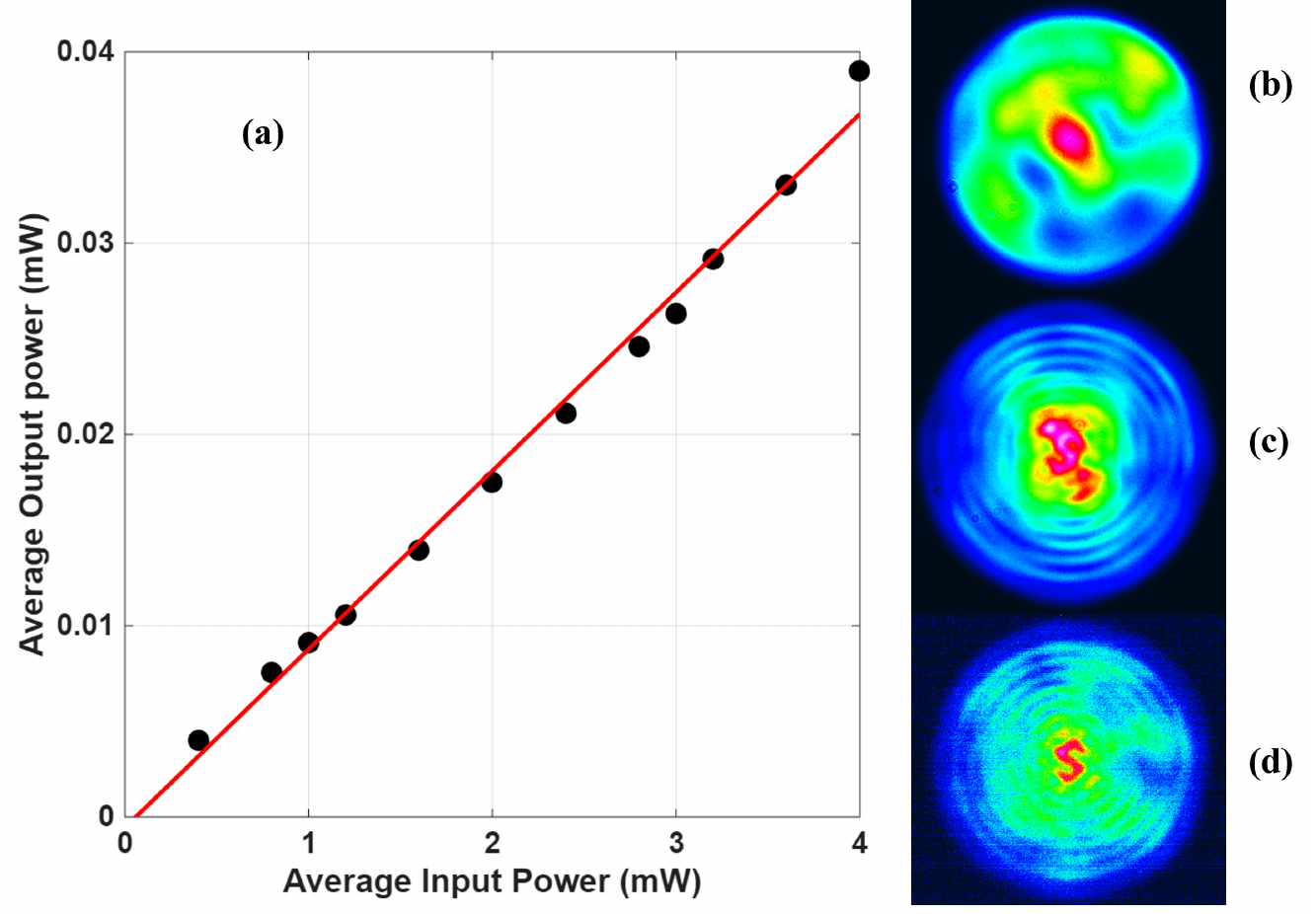}
    \caption{(a) The experimentally measured relationship between input and output power (black dots) accompanied by a linear fit (red solid line) for the SI MMF with acrylite coating. Near-field intensity profile captured during on-axis coupling is shown in (b), while (c) and (d) illustrate cladding coupling that produces Bessel-like LP mode rings, featuring speckle patterns at the center.}
    \label{fig:SI_tx_prof}
\end{figure}

To isolate the influence of the refractive-index profile on the generation and propagation of HOLG modes, we performed comparative experiments using SI MMF. By replicating the procedure used for the GRIN MMF—transitioning from on-axis Gaussian coupling to incremental spatial offsets—we sought to determine how the absence of a parabolic index landscape affects the resonant coupling. Under on-axis conditions (with the near-field intensity depicted in Fig.~\ref{fig:SI_tx_prof}b), the measured transmissivity was $82\%$, which aligns with the results obtained from the GRIN MMF. However, after introducing spatial offsets to generate structured light, transmissivity decreased to approximately $1\%$ (as illustrated in the input-output power relation in Fig.~\ref{fig:SI_tx_prof}a), resulting in the near-field intensity profile shown in Fig.~\ref{fig:SI_tx_prof}c. Alternative phase-matching conditions further produced the complex spatial distribution observed in Fig.~\ref{fig:SI_tx_prof}d.

\begin{figure}[pos=h]
    \centering
    \includegraphics[width=1.0\linewidth]{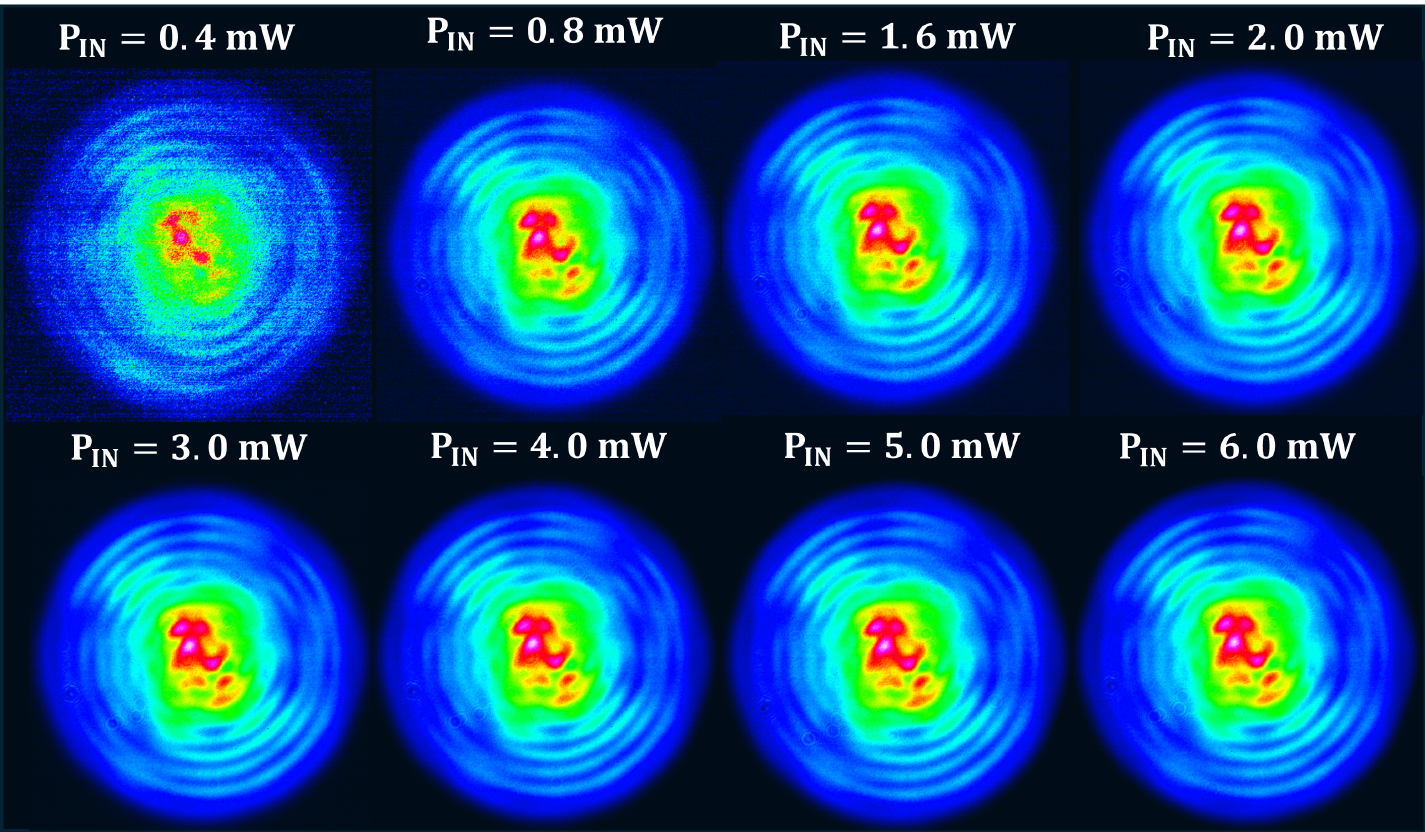}
    \caption{Near-field intensity profile from coated SI MMF demonstrates a Bessel-like LP mode with a central speckle beam.}
    \label{fig:SI_MMF_pwr}
\end{figure}

Near-field intensity profiles were monitored as a function of input power at the phase-matching point corresponding to Fig.~\ref{fig:SI_tx_prof}c. As illustrated in Fig.~\ref{fig:SI_MMF_pwr}, the SI MMF generates distinct concentric rings; however, unlike the GRIN MMF, the central region is dominated by a speckled output. Although these high-order structures exhibit significant morphological differences relative to the HOLG modes generated in the GRIN fibers, they remain invariant across the tested power range (0.4 mW to 6.0 mW). This power-independent behavior, combined with a stable transmissivity ranging from $0.86\%$ (at 1 mW) to $1.7\%$ (at 7 mW), definitively confirms that the observed beam shaping results from a purely linear optical process and excludes Kerr-induced mode mixing or nonlinear grating formation.

The significant divergence in the generated HOLG modes can be explained by examining the phase-matching conditions of the two fiber types. In a GRIN fiber, the parabolic refractive index profile gives rise to spatial eigenmodes (HOLG modes) with uniformly spaced propagation constants ($\beta$). As a result, the longitudinal phase-matching condition ($\Delta \beta \approx 0$) is exceptionally selective. The parabolic potential functions as a spatial filter, ensuring that resonant power coupling from the cladding exclusively excites a targeted single HOLG mode while effectively suppressing adjacent non-resonant modes. In contrast, the SI core is characterized by a flat, uniform refractive index, bordered by an abrupt step or a dip followed by a step (trench), resembling an infinite cylindrical square well. Here, the natural spatial eigenbasis transitions from LG to Bessel-like linearly polarized (LP) modes \cite{zhu2009generation}. More crucially, the propagation constants in an SI fiber are irregularly spaced. When the cladding light evanescently interacts with the SI core boundary, the phase-matching condition ceases to be singular. Instead, the cladding mode phase-matches with multiple core modes concurrently. This results in the dual structure of the observed SI output—outer rings and central speckle—which is dictated by the spatial overlap integral at the core-cladding boundary. 

According to Ivanov et al. \cite{ivanov2006cladding}, the coupling coefficient ($\kappa$) is influenced by the spatial overlap of the interacting mode fields at the interface. Since the optical power is primarily concentrated in the cladding, evanescent coupling tends to favor the excitation of the highest-order LP modes, similar to "whispering gallery" modes. In these modes, the Bessel-function intensity maxima lie at the outer edge of the core, maintaining the distinct concentric ring structure observed at the beam's periphery. However, unlike the precise harmonic filtering of the parabolic potential, the tunneled energy that moves toward the center of the SI core inevitably populates the lower- and mid-order LP modes. As these simultaneously excited modes propagate, they experience different phase delays due to intermodal dispersion. Upon exiting the fiber, their spatially overlapping electric fields coherently interfere, resulting in the static, multimodal interference pattern—the speckle—seen at the center of the beam (Fig.~\ref{fig:SI_MMF_pwr}). 
\begin{figure}[pos=h]
    \centering
    \includegraphics[width=1.0\linewidth]{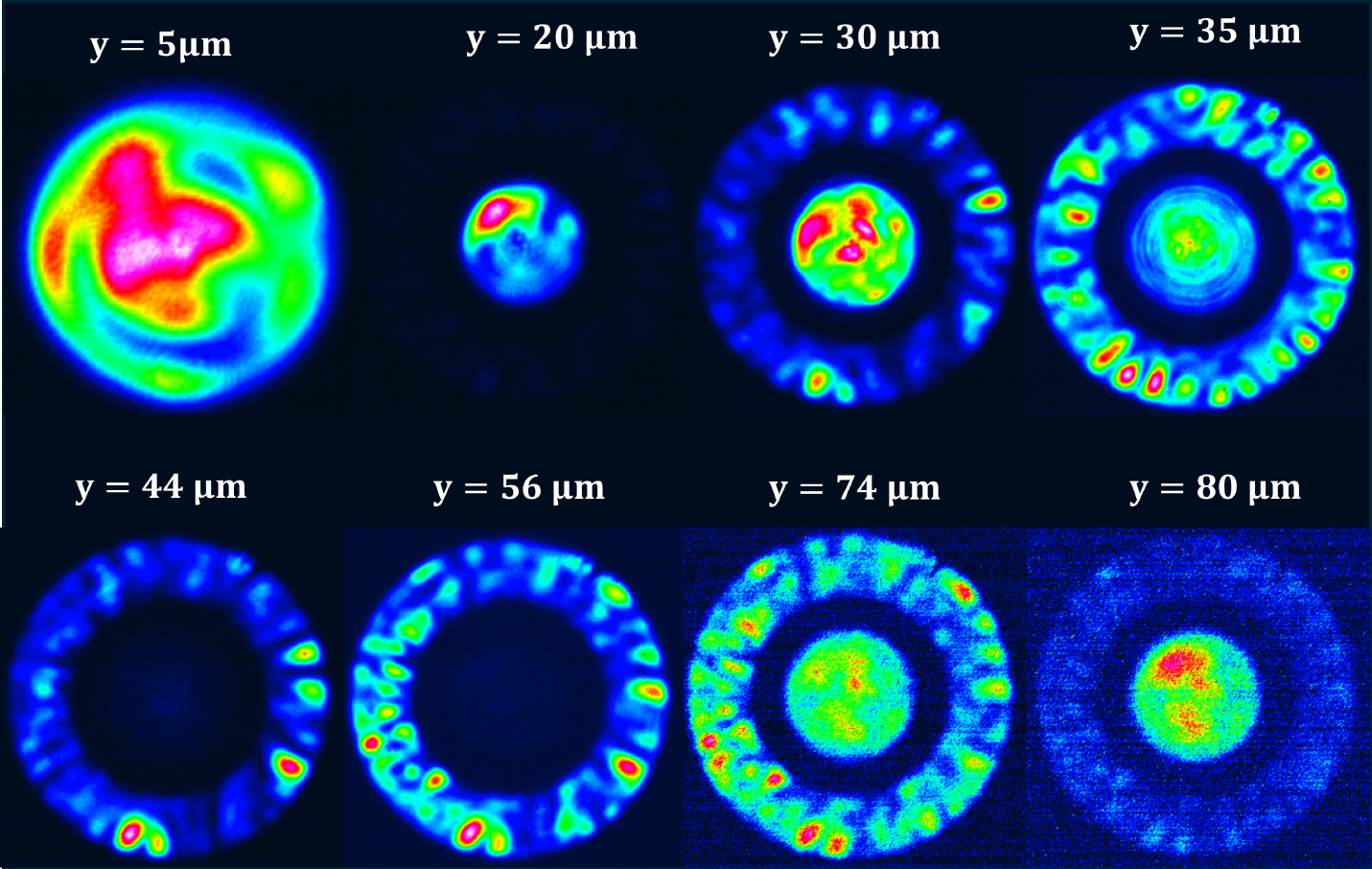}
    \caption{The near-field spatial evolution versus lateral input shift for the uncoated SI MMF shows a clear transformation from a speckled profile to a structured beam and beyond.}
    \label{fig:SI_uncoat_pwr}
\end{figure}

After establishing that the SI MMF produces a hybrid output, we conducted an additional experiment by removing the acrylic coating. By systematically translating the input Gaussian beam laterally—from direct core excitation to the cladding—we observed the progressive illumination of the cladding region, as illustrated in Fig.~\ref{fig:SI_uncoat_pwr}. A distinct dark ring separates this brightly illuminated cladding from the core, which serves as the optical signature of the depressed-index trench. Since the refractive index of this trench is lower than that of the core and the cladding ($n_{\text{trench}} < n_{\text{clad}} < n_{\text{core}}$), it functions as an exponential potential barrier. As a result, resonant energy coupling cannot occur through a direct geometric intersection; instead, the power within the cladding must experience frustrated total internal reflection at the high-contrast cladding-air boundary, allowing it to tunnel across the dark trench and populate the core modes. Remarkably, this trench-assisted tunneling explains the mechanism behind the hybridized state observed in the coated SI MMF (Fig.~\ref{fig:SI_uncoat_pwr}). 

\section{Conclusion}
In summary, this study presents a comprehensive framework for understanding and controlling the generation of HOLG beams from an input Gaussian beam through resonant cladding-to-core coupling in MMFs. By systematically examining the longitudinal and transversal dynamics of both picosecond and femtosecond pulses, we have shown that the order of the generated HOLG beam remains invariant across variations in power and propagation distance in GRIN MMF. Once resonant coupling occurs in the initial section of the fiber, the tunneled energy is directly projected onto the discrete eigenmodes supported by the core. Because the transverse modal structure is strictly governed by the waveguide boundary conditions, it propagates forward as a stable eigenstate. Consequently, while the temporal walk-off inherent to the ultrafast regime limits the effective interaction length and reduces the absolute conversion efficiency for ultrashort pulses, it does not modify the transverse modal order of the generated beam.

Through a rigorous comparative analysis, we further demonstrated that the transverse refractive index landscape acts as the ultimate spatial filter, governed by the coupled-mode overlap integral. In a GRIN MMF, the parabolic potential behaves as a strict harmonic oscillator, effectively distilling evanescently coupled energy into pristine, topologically protected HOLG modes. In contrast, a standard step-index configuration lacks this harmonic isolation, leading to a hybridized beam with a Bessel-like LP mode and a central speckle. Finally, power-scaling and spectral-broadening analyses unequivocally demonstrate that this beam-shaping mechanism is fundamentally a linear optical process. Although the intrinsic Kerr nonlinearity of the silica medium results in SPM at high peak intensities—slightly altering the longitudinal phase-matching condition and leading to a sublinear power response—the order of the HOLG modes remains entirely protected. As a result, this boundary-enforced core-cladding resonant coupling provides a highly deterministic, linear, and power-scalable platform for generating tailored HOLG modes.


\section*{Declaration of competing interest}
The authors declare that they have no known competing financial interests or personal relationships that could have appeared to influence the work reported in this paper.

\section*{Data availability}
Data underlying the results presented in this paper are not publicly available at this time but may be obtained from the authors upon reasonable request.

\section*{Funding} 
The authors acknowledge the support from the European Union, Next Generation EU, through the Project PRIN 2022 “Spiral and Focused Electromagnetic fields” (SAFE) of the Italian Ministry of University and Research (MUR), under Grant 2022ESAC3K.

%\appendix
%\section{Example Appendix Section}
%\label{app1}




\bibliographystyle{elsarticle-num}
\bibliography{sample}

\end{document}